# THE IMPACT OF COUNTDOWN CLOCKS ON SUBWAY RIDERSHIP IN NEW YORK CITY

**Zhengbo Zou[1], Di Sha[2]**

[1]PhD Candidate, Civil and Urban Engineering, New York University
6 MetroTech, New York, 11201
Email: zz1658@nyu.edu
[2]PhD Candidate, Civil and Urban Engineering, New York University
6 MetroTech, New York, 11201
Email: ds5317@nyu.edu

ABSTRACT

Protecting the passengers' safety and increasing ridership are two never ending pursuits of public transit agencies. One of the proposed methods to achieve both goals for subway service is to implement real-time train arriving countdown clocks in subway stations. Metropolitan Transportation Athoriy (MTA) of New York City (NYC) chose to install such countdown clocks in their stations starting from 2007 on a selection of subway lines. Due to the recent development of Bluetooth Beacon technology, the MTA could now install countdown clocks and train trackers in a non-intrusive manner with much faster speed. As a result, the MTA is aiming to install countdown clocks in every subway station on every line by the end of 2017. However, with such an aggressive plan, the impact of countdown clocks on subway ridership has not been fully studied. This paper proposes using Panel Regression methods, specifically, Random Effect (RE) model and Fixed Effect (FE) model to quantify the impact of countdown clocks on subway ridership. Machine Learning methods, namely Random Forest (RF) with AdaBoost and Decision Tree (DT) Regression, are also used as alternative data-driven approaches for the FE/RE model. The results show that for the G line service, which runs between Brooklyn and Queens, the introduction of countdown clocks could increase weekly ridership by about 1783 per station. The study also found that the machine learning methods provide better accuracy in predicting the ridership than RE/FE models.

# INTRODUCTION

Public transportation plays a vital role in metropolitans such as New York City (NYC). In the year 2016, the Metropolitan Transportation Athoriy (MTA) reported a total subway ridership of 1.757 billion, ranking the first in the U.S., and the 7th around the globe (MTA 2016). Public transit (1) lowers transportation cost for consumers (Litman 2015); (2) decreases traffic congestion in large cities (Arnott 1994); (3) has the potential to improve health and fitness for travelers (Litman 2003, Zou and Ergan 2018); and (4) creates vibrant and wealthy communities around the transit lines (Vuchic 2017).

Although with various advantages, issues remain in multiple areas of public transit systems, such as reliabitlity, safety and rider satisfactory (Walker 2011). Safety is one of the most important goals of public transit agencies, but subway fatalities are still hauting transit agencies. Between Janurary 1st 2003 to May 31st 2007, the total fatalities involving MTA subway trains are 211 (Lin 2009). Another persuit for public transit agencies is rider satisfaction. According to the 2015 MTA satisfaction survey (MTA 2015), the overall satisfaction rate of MTA is 83%, and the on-time performance is only 74%. These issues urge change and improvement from transit agencies. One of the proposed solutions to improve both rider satisfaction and rider safety is to provide real-time information, such as real-time train arrival prediction, to the costumers in various forms.

With the development of information technology and consumer electronics, real time information could be delivered to the passengers in many forms (Du et al., 2018, Kasireddy et al., 2016, Zou et al., 2018). For example, the MTA started to install countdown clocks on the numbered line train stations starting from 2007 (MTA 2009). Such physical signs in the subway stations could provide much-needed train arrival prediction to a large audience without any direct investment from the riders. The prevailence of handheld electronic devices brings the real-time information to the fingertips of transit costumers. For example, MTA started to gradually roll out the real-time bus arrival information to bus riders from 2010 (MTA 2010). The information could be accessed online, so any devices that supports web-based applications could benefit from the information.

The implementation of real-time information platforms (i.e., countdown clocks and web-based apps) requires significant investment from public transit agencies. However, the impact of such real-time information on subway ridership has not been fully studied. This paper investigates the impact of countdown clocks on subway ridership by analyzing ridership variation before and after the installation of countdown clocks in subway stations. The study also takes into account of other factors that might contribute to the changes in subway ridership, such as temperature, precipitation, and unemployment rate. The study uses Panel Regression models, namely Random/Fixed Effect Model, and Machine Learning algorithms, namely Decision Tree Regression and Random Forest with Adaboost. A comparison of various methods was conducted, and the result shows that for the G line running between Brooklyn and Queens, the ridership increases around 1783 weekly per station, and the Machine Learning techniques provide higher prediction accuracy than Panel Regression methods.

The paper is organized as follows. First, a literature review on the impact of real-time information on transit riders is provided. Next, the data and methodology are discussed. Finally, results are shown, followed by the conclusion and future work.

# LITERATURE REVIEW

The provision of real-time arrival information could be separated to two categories (i.e., physical countdown clocks showing next train arrival time, and web-based applications providing train schedule and train arrival time). Previous research (Tang 2012, Brakewood 2015, Ferris 2011, Chow 2014) suggests that real-time information have the following benefits to transit customers:

1) Reduced the perceived wait time, and in turn reduced anxiety.
2) Improved the perception of transit agencies and improved satisfaction rates.
3) Improved ridership for various transit modes.

**Ridership Impacts of Real-time Information from Web-based Applications**

There has been ample research on the impact of real-time travel information provided through online websites and phone applications. Most notably, (Ferris and Watkins 2010, 2011), (Tang 2012), and (Brakewood 2014, 2015) studied the impact of real-time information on various transportation means in many cities in the U.S. Ferris and Watkins conducted online and in station surveys. The results showed that the provision of real-time information increased trips per week reported by the passengers, with a larger gain from non-commute trips. Tang studied the impact of real-time bus information system implemented in Chicago. The analysis focused on the average weekday bus ridership vairations from 2002 to 2010, with real-time information introduced in May 2006. Tang found that there was a modest impact of real-time information, attributing an increase of 126 trips per weekday. Brakewood studied the impact of real-time bus information, using NYC bus system as a testbed. The result showed an average increase of 118 unlinked trips per route per weekday. The study also found that the improvement in ridership is largely attributed to large routes with high revenue miles of service.

**Ridership Impacts of Real-time Information from Phyiscal Countdown Clocks**

Previous research on real-time countdown clocks in stations often looks into the rider satisfactory and perceived wait time effect (Hickman 1995, Mishalani 2000). The ridership effect of countdown clocks in subway or bus stations has not been fully studied. One possible reason is that countdown clocks were installed starting in the late 1990s and the early 2000s. In the mean time, the ridership data was not abundantly available. As a result, the studies were mostly conducted based on survey results, which provided little implications on ridership. The studies that investigated the ridership effects from countdown clocks found minor improvements in ridership due to the introduction of real-time clocks (Chow 2014).

In conclusion, previous research on real-time information usually focused on the web-based applications. The studies often found that the impact on ridership to be "moderate" or "minor". The studies that focused on physical countdown clocks showed limited results on ridership because the research mostly focused on customer satisfaction and perceived wait time. One of the challenges to infer ridership changes from real-time information provision is that multiple factors could attribute to the ridership variation. Omission of such factors would create models that do not reflect real-world changes in ridership.

# DATA INTRODUCTION

The MTA of New York City first introduced the real-time arrival information signage (i.e., countdown clocks) on the numbered lines (i.e., line 1 through line 7) starting from 2007 (MTA 2009). The countdown clocks are electronic signs in subway stations that let customers know when the next train is coming, as shown in Figure 1. Aside from next train arrival information, countdown clocks are used to (2) provide emergency information; (3) show train service type (i.e., local or express); and (4) eliminate the temptation to peer down platforms looking for trains (MTA 2017).

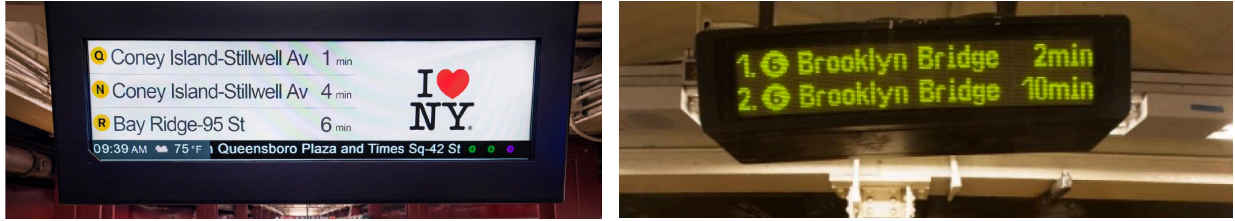

**FIGURE 1. Real-time Countdown Clocks Showing Train Arrival Times in Subway Stations.**

Since the first introduction of countdown clocks in 2007 on the numbered lines, the MTA has installed more than 100 clocks in stations. The process was slow and costly. Recent development of low cost Bluetooth Beacon technology has provided a non-intrusive way to install a new type of countdown clocks, while significant reduced the cost. As a result, the MTA started to pick up the pace of installing the countdown clocks on the lettered lines stations. The goal is to install countdown clocks in every station in NYC by the end of 2017 (MTA 2017).

This paper uses the gradual installation of the new countdown clocks in the lettered lines as a natural experiment, and investigates the impact of such clocks on the subway ridership, while controlling other factors such as temperature, precipitation and umemployment rate. The first completed installation of new countdown clocks using Bluetooth Beacon technology was on August 31$^{st}$ 2017, for all stations of the E line and G line. The author used this date as the separation point for a before/after study on both the E and G line.

The ridership data was collected from the New York State Open Data (New York State 2017). The ridership data consists of number of MetroCard swipes made each week by customers entering each station of the New York City Subway, PTAH, AirTrain JFK and Roosevelt Island Tram. The data is organized by the type of MetroCard used (i.e., 30 days unlimited or full fare one time card). The dataset was first filtered to only consisting E and G line data. The author then grouped card swipe counts by card type. Finally, card swipe counts were separated by subway lines, meaning each line has its own table showing total number of swipes per week, for each station. An example of G line card swipe counts is shown in Table 1.

**TABLE 1. Fare Card Swipe Count for NYC Subway Stations for the G Line.**

| Date | Station Name | Total Count |
|---|---|---|
| 6/3/17 | Line_G 15 ST-PROSPECT | 37694 |
| 6/3/17 | Line_G 4 AVE | 87556 |
| 6/3/17 | Line_G 7 AV-PARK SLOPE | 67972 |
| 6/3/17 | Line_G BEDFORD/NOSTRAN | 50880 |

| | | | | | | |
|---|---|---|---|---|---|---|
| 6/3/17 | | | | Line_G BERGEN ST | 67994 | |

Aside from the subway ridership data, other data sources were introduced to account for ridership changes. Previous research shows that external factors for ridership changes include weather, unemployment rate, transit fare, availability of other transit services such as CitiBike, and natural disasters such as hurricane Sandy (Brakewood 2015, Tang 2012). Based on the previous research and the availability of data, this paper included weather data from the National Weather Service Forcast Office of New York City (NWSFO 2017), and unemployment data from the U.S. Department of Labor (DOL 2017). The weather data was collected at the NYC Central Park area. Daily weather data was exported. The data consists daily high/low/average temperatures in Farenheights and precipitation in inches. For convenience of the modeling process and computational speed, the author grouped the temperature into three categories, i.e., low (below 40), moderate (40 to 75) and high (above 75). The precipitation data was included as is. Unemployment data was collected from the Bureau of Labor Statistics, and the data was provided as a percentage by month. For example, the unemployment for November 2017 is documented as 4.1%. Finally, a dummy variable called "HasClock" was introduced to represent the existence of countdown clocks. So, for the dates before and after August 31$^{st}$, the value would be zero and one respectively. Once all data sources were merged, tables of subway ridership with external factors could be created for each line. An example of the G line is shown in Table 2.

**TABLE 2. Ridership Data with External Factors for the G Line.**

| Date | Precipitation | Category | Unemploy | Station Name | Total Count | Has Clock |
|---|---|---|---|---|---|---|
| 6/3/17 | 0.0257 | moderate | 4.4 | Line_G 15 ST-PROSPECT | 37694 | 0 |
| 6/3/17 | 0.0257 | moderate | 4.4 | Line_G 4 AVE | 87556 | 0 |
| 6/3/17 | 0.0257 | moderate | 4.4 | Line_G 7 AV-PARK SLOPE | 67972 | 0 |
| 6/3/17 | 0.0257 | moderate | 4.4 | Line_G BEDFORD/NOSTRAN | 50880 | 0 |
| 6/3/17 | 0.0257 | moderate | 4.4 | Line_G BERGEN ST | 67994 | 0 |

**METHODOLOGY**

This paper quantifies the impact of real-time countdown clocks on subway ridership using two types of methods, Panel Regression Models and Machine Learning techniques. Panel Regression Models allow the author to explicitly quantify each factor's effect on ridership, while the Machine Learning methods provide data-driven approaches to predict the ridership changes based on existing data.

As shown in Figure 2, a closer look into the monthly ridership for the MTA Bus and Subway, would reveal that ridership has a seasonal pattern. Ridership reaches the highest levels during spring and fall months, while decreases to the lowest levels in summer and winter months. Moving average for bus and subway, shown as the solid lines in Figure 2, indicates an upward trend for subway ridership, and a downward trend for buses. These time related effects should be taken into account when building the models. As a result, in the Panel Regression Models, Fixed

Effect Model (FE) was considered. For accuracy of the model, a Random Effect Model (RE) was also used for comparison. Hausman test was done to determine the significance of difference between these two models. For Machine Learning methods, Random Forest (RF) with Adaboost was used because of the high prediction accuracy this method has shown in regression problems. Regular Decision Tree Regression (DT) was used as a comparision for the RF method.

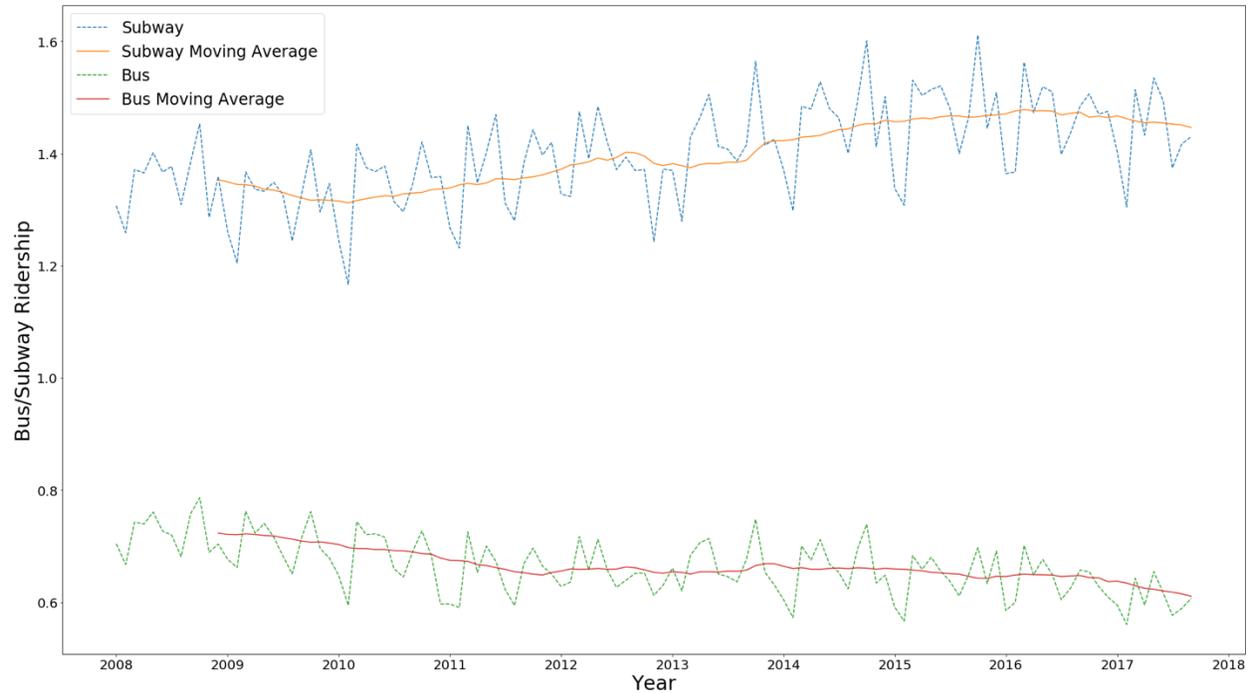

**FIGURE 2. NYC Bus and Subway Ridership by Time**

**Panel Regression Models**

Two types of Panel Regression Models were used in this study, namely Fixed Effect (FE) model and Random Effect (RE) model. FE model has shown great value when the number of factors are high, and correlation exists within factors in the panel data. On the other hand, RE models assume that factors within panel data are independent. The independent assumption can be an over-simplification for plenty of datasets, hence giving sub-optimal results when comparing to FE models. To examine the correlation among factors, this study conducted the Hausman test, with the null hypothesis being that the FE model and RE models are statistically similar. If the p-value came larger the 0.05, the null hypothesis would be accepted. In the case where FE and RE models are not statistically siginifantly different, RE models should be chosen because of its computational efficiency.

*Fixed Effect Models*

Fixed Effect Models are built on Ordinary Least Square (OLS) Regression:

$$Ridership_{it} = \alpha + \beta x_{it} + \varepsilon_{it} \quad (1)$$

Where $i \in (1, \dots, N)$ represents distinct subway stations, and $T \in (1, \dots, T)$ represents time. Ridership is the dependent variable, or target value, while the factors such as temperature and precipitation are independent variables (i.e., $x$ in equation 1). $\alpha$ in equation (1) represents the intercept of the model, and $\beta$ is the vector of estimated coefficients. The main assumption for OLS regression is that the error term $\varepsilon_{it}$ is assumed to be independent identically distributed (IID) random variable. The IID assumption does not hold if there are correlations within factors.

FE models vary from the OLS model by separating the error term $\varepsilon_{it}$ into two distinct terms: individual effect $\mu_i$ and idiosyncratic error $v_{it}$. The FE model further considers the correlation among factors by introducing individual means of factors $\bar{x}_i$. The introduction of individual means decreases the degree of freedom of the model, but gives consistant results when correlation exists among factors. FE models is described using equation (2).

$$Ridership_{it} = \overline{Ridership_{it}} + \beta(x_{it} - \bar{x}_{it}) + v_{it} \quad (2)$$

*Random Effect Models*

RE models differ from FE models by assuming factors $x_i$ are uncorrelated. This assumption does not eliminate the degree of freedom like the FE model, hence computational efficient. However, the over-simplification could provide inconsistent models. In RE models, intercepts could be on an individual level, meaning $\alpha$ in OLS models is now $\alpha_i = \alpha + \mu_i$. RE model is described in equation (3).

$$Ridership_{it} = \alpha_i + \beta x_{it} + v_{it} \quad (3)$$

To examine if the independent assumption holds in RE models, Hausman test (Hausman 1981) could be conducted. Hausman test assumes the similarity between FE model and RE model are high. If the p-value of Hausman test is less than 0.05, the null hypothesis should be rejected. Otherwise, the FE model and RE model would be proven similar, and RE model should be adopted for computational efficiency.

**Machine Learning Methods**

Two Machine Learning methods were used in this study, namely Random Forest (RF) with Adaboost, and Decision Tree (DT) Regression. The use of RF models is due to its great performance on supervised learning. (Caruana 2006) compared ten most widely used supervised learning techniques (i.e., machine learning methods with labeled training set) on empirical datasets, and found that Random Forest algorithm being the second-best performer, only after Boosted Trees with tuned parameters. Admittedly, the development of Deep Learning in recent years may have shadowed the performance of Random Forest, but Deep Learning requires comparibly larger datasets, and longer time to train. As a result, for this study, Random Forest algorithm was selected as the prediction algorithm. Adaboost is also implemented in the RF algorithm, because it generally provides even better prediction results than basic RF algorithm. As a comparison, regular Decision Tree (DT) Regression algorithm was used as a baseline for prediction.

*Random Forest with Adaboost*

Random Forest (RF) algorithm is one of the ensemble algorithms. Ensemble algorithms generate many base models (classifiers or regressors), and then aggregate the results from individual base models (Liaw 2002). In theory, the base classifiers or regressors could be any algorithm. In practice, Decition Tree is often used as the base classifier or regressor. Random Forest is built on the concept of boosting, which creates base models in order. The successive base model would look at the previous base model's performance, and improve its own performance based on the previous base model's errors. Adaboost is one of the boosting techniques. Adaboost could be illustrated in Figure 3, where the whole training dataset consists of various training samples (i.e., black, blue, red and green data points). The first base model (Base Model #1) randomly selects a subset of the training set (blue and red training points), and produces a regression model. For Base Model #1, the result may show that it accurately predictes the value of blue points, but not the red points. As a result, for the second base model, the training sample now includes not only the randomly selected training points (black points), but also previously inaccurately predicted points (red points). Base Model #2 then produces another result (result #2). Notice this time, the regression model would perform better on the previously inaccurately predicted red points, but it might suffer on some other training data, such as the black points. The following base models, like Base Model #2, take into account of the errors made in the previous models, and try to produce a better model that does better on those errors. After N iterations, the algorithm produces an aggregated result from all individual base models. The process is called Adaboost, or adaptive boost.

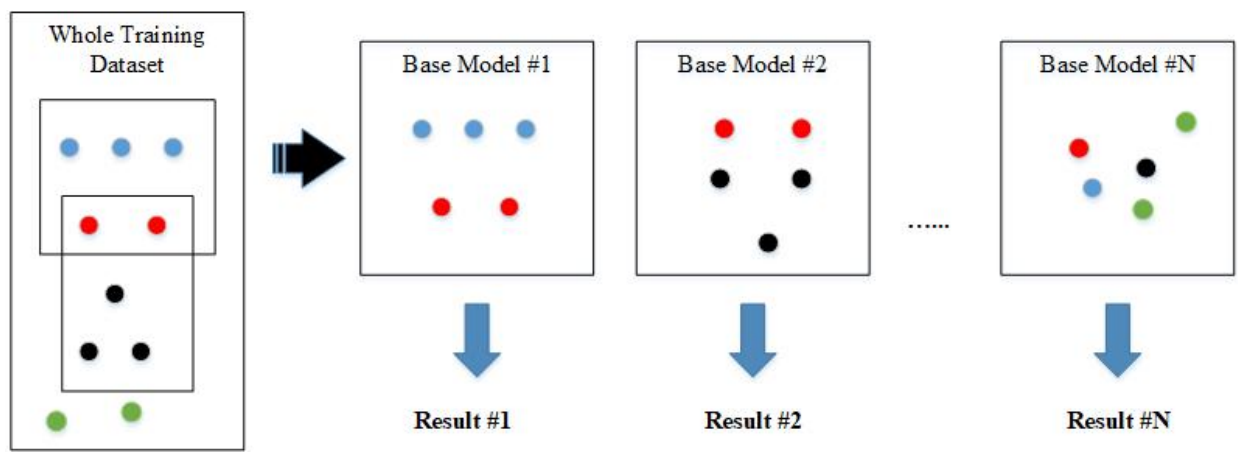

FIGURE 3. Boosting Machanism of Random Forest Algorithm

Random Forest adds another layer of randomness on top of boosting. The algorithm randomly selects a subset of features, each time a base model is created. For example, the algorithm might choose precipitation and temperature to create the first base model, and choose precipitation and unemployment to create the second base model. In this study, RF algorithm was adopted to predict the ridership based on the training data collected, as previously shown in Table 2.

*Decision Tree Regression*

In the RF algorithm, the base model is often selected as the regular Decision Tree (DT) Regression. As a result, this paper included the DT algorithm as a baseline comparison to examine if the addition of boosting and random selection of training features created any improvement on the overall prediction model.

The DT algorithm is created based on a Tree Structure, as shown in Figure 4. The decision nodes are defined as the nodes that could be futher separated, while leaf nodes are defined as the ones that could not be separated anymore. In Figure 4, any nodes that does not represent the target value (Ridership) are decision nodes. Branches in regression trees are carrying wights that represent the confidence level of the choices made by the branches. The final regression value is a weighted aggregated value from all leaf nodes.

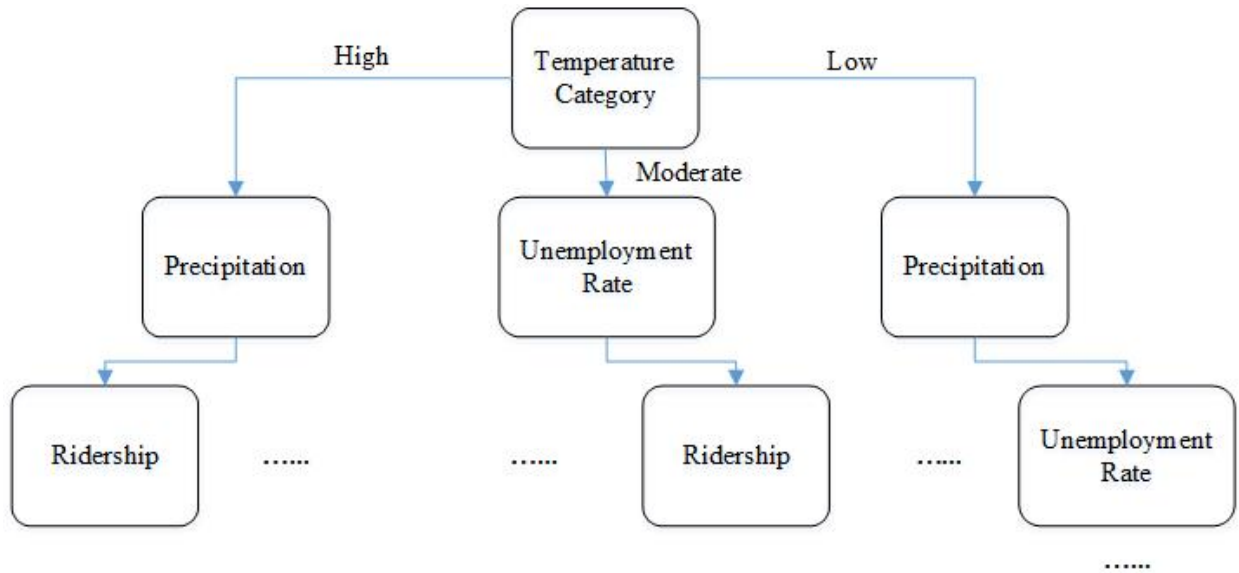

**FIGURE 4. Decision Tree Regression.**

Decision Trees seperate on the node (i.e., feature from training data) produces the least mean square error after the seperation. This type of separation technique is called ID3 algorithm (Quinlan 1986), and it is the basis of many advanced separation methods. As in this study, the separation method adopted is ID3.

**RESULT AND DISCUSSION**

The result is presented as follows. First, the result of FE and RE models are shown to provide insights on individual factors' coeffients. Next, the result of RF and DT algorithms are provided, showing the parameter tuning process for the algorithms. Finally, Panel Regression models and Machine Learning methods are compared based on the bias and variance of each methods.

**Panel Regression Models**

The study was conducted on a station level for E and G line from June 1st 2017 to November 30th 2017. The result of Panel Regression models is shown in Table 3. It should be noted that the coefficients estimated by RE and FE models are identical in this study. A futher Hausman test also returned a p-value of 1, meaning the two models are identical when using the provided panel data. As a result, the RE model was selected for futher predictions due to its computational efficiency.

The coefficients could be interpreted as the impact on ridership from each factor, before and after the installation of countdown clocks. For example, on average, a ridership improvement of 1783 per week per station would occur on the G line service, due to the introduction of countdown clocks. It should be noted that the "HasClock" dummy variable has a p-value slightly lower than 0.05 in the G line analysis and a large p-value in the E line analysis, meaning the introduction of countdown clocks in E and G line do not cause significant impacts on ridership. Both analysis for E and G line show that the precipitation was selected as the only statistically significant factor. As of the goodness of fit for both G and E line, RE and FE models reported low R-Squred values, which implies that the RE/FE models could not explain the ridership impact caused by the factors in the panel data. Altough the result of FE and RE models are not optimal, the modeling process is valuable. The author used RE/FE models to explicitly list the factors that might be influencing subway ridership, and the causal relationships are of great importance for future research on the subject.

**TABLE 3. FE/RE Model Results for G and E Line.**

|  | G Line | | | | E Line | | | |
|---|---|---|---|---|---|---|---|---|
|  | RE Model | | FE Model | | RE Model | | FE Model | |
|  | Coefficient | p-value | Coefficient | p-value | Coefficient | p-value | Coefficient | p-value |
| Intercept | 29713.61 | 0.08 | - | - | 142004.98 | 0.06 | - | - |
| Precipitation | -6232.39 | **1.78e-5** | -6232.39 | **1.78e-5** | -20733.40 | **1.78e-4** | -20733.40 | **1.78e-4** |
| Temperature Category | -677.45 | 0.09 | -677.45 | 0.09 | -936.94 | 0.54 | -936.94 | 0.54 |
| Unemplpyment Rate | 3996.05 | 0.29 | 3996.05 | 0.29 | 13810.86 | 0.34 | 13810.86 | 0.34 |
| Has Clock | 1783.37 | 0.06 | 1783.37 | 0.06 | 3608.62 | 0.32 | 3608.62 | 0.32 |
| R Squared | 0.09 | | 0.09 | | 0.03 | | 0.03 | |

**Machine Learning Algorithms**

To achieve the optimal performance of RF algorithm and DT regression algorithm, vital parameters need to be tuned. For the RF algorithm, important parameters are (1) the depth of the tree, meaning the number of seperations for each base model (implemented using regular decision tree), and (2) the number of estimators, meaning the number of base models created. For the DT regression algorithm, the important parameter is the depth of the tree. For the accuracy of RF and DT algorithms, cross validation technique was used to iteratively generate a training and testing subsets to tune the parameters. This study used a standard 80-20 split for training-testing seperation.

During the tunig of depth of tree and number of estimators for RF algorithm, mean square error (MSE) and residual sum of squres (RSS) were used as measurements for variance and bias for the algorithm. A low MSE would suggest a low variance of the algorithm, while a low RSS would suggest a low bias. The optimal result would be a model with both low bias (RSS) and low variance (MSE). The results for RF algorithm are shown in Figure 5 and 6 for G and E line.

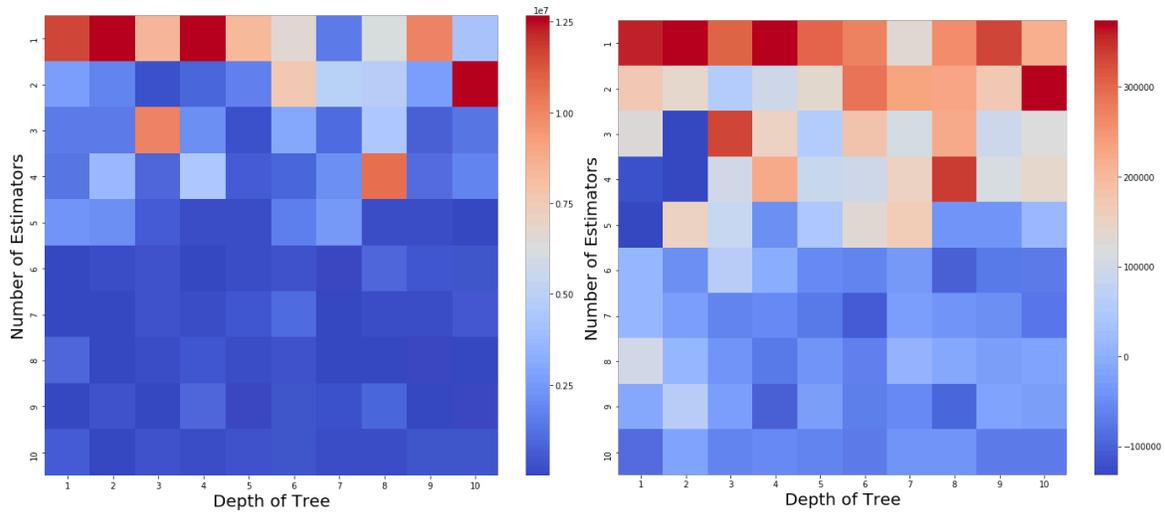
**FIGURE 5. MSE (left) and RSS (right) for RF algorithm of G line.**

As shown in Figure 5, the best MSE and RSS results are clustered on the lower right coner, where the value of MSE and RSS are the lowest, but the number of estimator and depth of tree are the highest. However, such as combination of parameters would almost certainly cause overfitting. In fact, the choice of parameters should also obey occam's razor: when accuracy levels produced by different models are the same, the model with the lowest level of complexity should be selected as the final model. As a result, the occam's razor rule pushes the parameter selection toward the upper left corner. These two contradicting rules produced two parameter sets in the G line RF algorithm analysis. One set of parameter is: number of estimators being 2, depth of tree being 3. The other set of parameter is: number of estimators being 3, depth of tree being 8. To select a final model from these two sets of parameters, R-Squared was used as a tie-breaking measurement. The set of parameter that achieves the highest R-Squared value (3, 8) was selected as the final parameters for the RF algorithm.

Similar process of parameter tuning was conducted for the E line. The MSE and RSS are shown in Figure 6. The selected parameters for RF algorithm for the E line were: number of estimator being 4, and the depth of tree being 7.

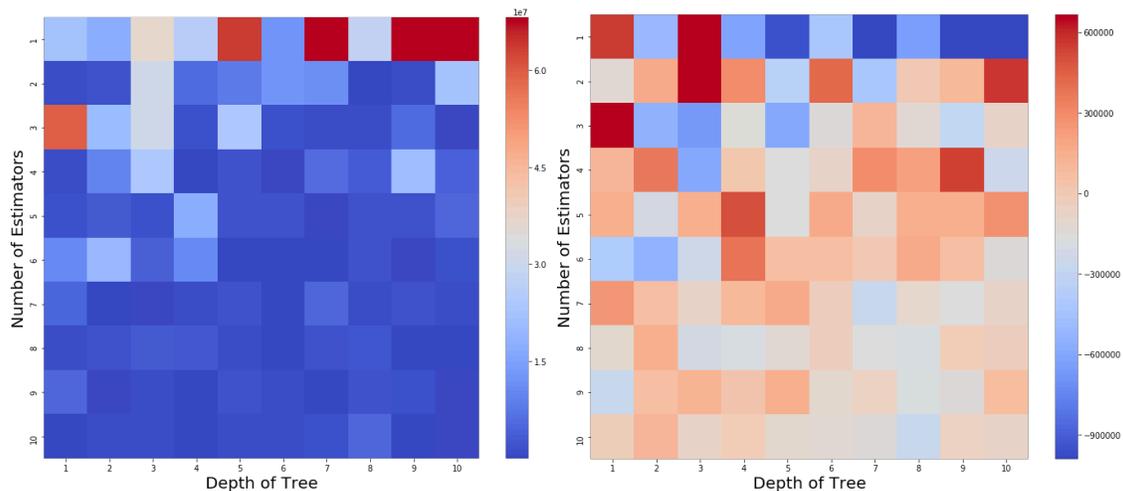
**FIGURE 6. MSE (left) and RSS (right) for RF algorithm of E line.**

To fairly compare the RF algorithm with regular DT regression, the same depth of tree was used in the DT regression. The measurement used was R-Squred value. As shown in Table 4, the DT regression actually produced better R-Squared values than the RF algorithm, meaning the boosting effect was not significant in the RF algorithm. As a result, the regular DT regression was selected as the Machine Learning prediction model.

**TABLE 4. R-Squared Value for DT and RF Algorithms.**

|  | G line | | E line | |
| --- | --- | --- | --- | --- |
|  | DT | RF | DT | RF |
| R-Squared | 0.95 | 0.90 | 0.99 | 0.98 |

**Prediction Comparison**

The prediction evaluation between Panel Regression models and Machine Learning techniques was done by comparing the accuracy of prediction produced by each method. As metioned in the result section, the Panel Regression model selected was Random Effect model, and the Machine Learning method selected was Decision Tree Regression. Just as a comparison, Random Foreset result was also added in the graph. The comparison result for both the G and E line are shown in Figure 7. The x-axis in the figure is the true ridership value, and y-axis is the predicted ridership. A good prediction model would produce points that evenly and closely surround the solid 45-degree line. It could be seen from the figure, that RE model produced a horizontal distributed prediction result, meaning the model predicts similar ridership no matter what the real ridership might be. This also echoes the low R-Squared value from RE and FE models. The RF and DT predictions generally fall evenly on both sides of the solid line, and closely resembel the true ridership.

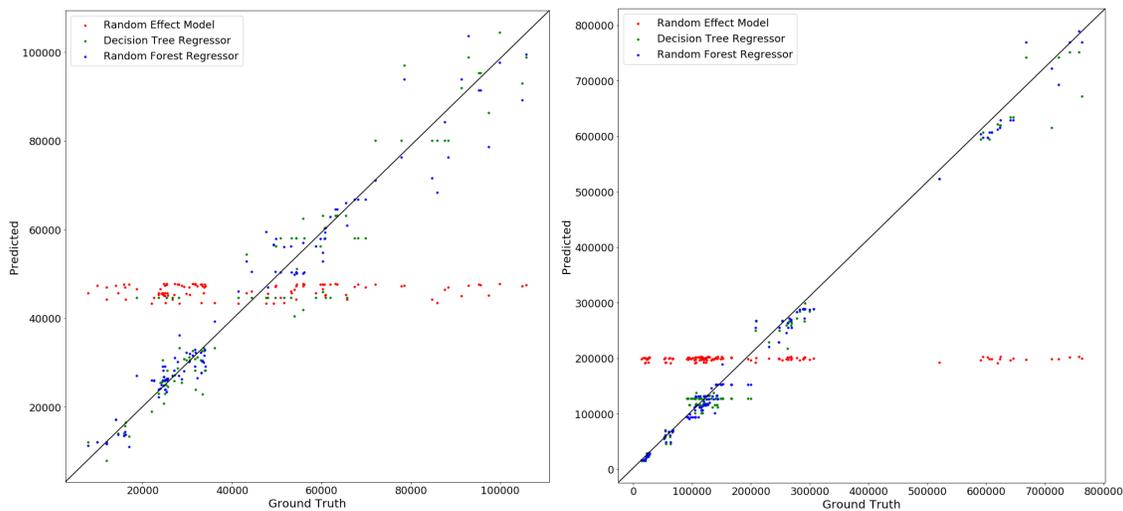

**FIGURE 7. Prediction Comparison among RE Model, RF Algorithm and DT Regression for G (left) and E (right) Line.**

**CONCLUSION**

This paper investigates the impact of real-time subway arrival information, specifically the information shown using countdown clocks in subway staitions, on ridership in New York City subway E and G line. Panel Regression models were used to explicitly list the possible factors that might influence the ridership, including average temperature, precipitation, unemployment rate, and whether the countdown clocks are installed. Machine Learning techniques were also considered as alternative data-driven approaches to predict the impact of countdown clocks on subway ridership. The result shows a moderate increase on subway ridership due to the introduction of countdown clocks: 1783 more people per station per week for the G line. The result also reveals that the Machine Learning approaches provide more accurate predictions comparing to the Panel Regression models. Finally, when considering prediction of subway ridership with panel data in the future, Random Forest with Adaboost and Decision Tree Regression could provide satisfactory results.

**LIMITATION AND FUTURE WORK**

The main limitation of this paper is the poor performance by the Panel Regression models. The reasons could be concluded into three aspects:

1. *The analysis period is too short (6 months with 3 before the installation of countdown clocks and 3 after).* Short analysis period means small dataset for the Panel Regression models to estimate. With larger pool of data, the performance is likely to improve.
2. *The factors considered in Panel Regression are not comprehensive.* This study only considered temperature, precipitation, and unemployment rate as additional factors other than the availability of countdown clocks. The result shows only precipitation has a significant impact on subway ridership. Previous research found more factors might have impact on ridership, such as the availability of alternative transit mode, fare price, and gas price. The inclusion of additional factors would certainly change the estimation model, and create more accurate estimates for the coefficients.
3. *The assumption of linearity of Panel Regression model may not hold.* The Panel Regression models used in this study are variations of a linear regression model. If the data does not obey the linearity given all factors considered, the model would never produce satisfactory results.

The problem of analysis period could be solved when data is abundantly available. The MTA has already installed countdown clocks in more than 500 stations in NYC, and the rest are being installed. The problem of additional factors could easiliy be solved by adding more data sources into the estimation. The problem of linearity is the hardest to solve. The only way to amend this is to adopt other types of models that do not only produce linear models. As an example, the Machine Learning algorithms used in this study, Random Forest and Decision Tree could produce non-linear models, making them better fit for the data, and they performed well in prediction comparisons.